\documentclass[useAMS,usenatbib, usegraphicx]{mn2e}
\usepackage{epsfig}
\usepackage{graphicx}
\usepackage{amsmath}
\usepackage{amssymb}
\usepackage{natbib}
\newcommand{\kms}{\ensuremath{{\rm km\,s}^{-1}}}

\newcommand{\msun}{\ensuremath{M_{\odot}}}

\newcommand{\infinity}{{\infty}}
\newcommand{\apj}{ApJ}
\newcommand{\apjl}{ApJ}
\newcommand{\mnras}{MNRAS}

\newcommand{\apjs}{ApJS}
\newcommand{\nat}{Nat}
\newcommand{\araa}{ARAA}

\newcommand{\physrep}{Physics Reports}

\newcommand{\HI}{H\,\textsc{i}}

\begin{document}

\title[Probing Reionization with MW Satellites]{Probing the Epoch of Reionization with Milky Way Satellites}

\author[Mu{\~n}oz, Madau, Loeb \& Diemand]{
Joseph A.\ Mu{\~n}oz$^1$\thanks{E-mail:jamunoz@cfa.harvard.edu}, 
Piero Madau$^2$,
Abraham Loeb$^1$, and 
J\"{u}rg Diemand$^{2,3}$\\
	$^1$Harvard-Smithsonian Center for Astrophysics, 
			60 Garden St., MS 10, Cambridge, MA 02138, USA\\
	$^2$Department of Astronomy \& Astrophysics, University of California Santa Cruz, 
			1156 High Street, Santa Cruz, CA 95064, USA\\
	$^3$Hubble Fellow
}

\maketitle

\begin{abstract}

While the connection between high-redshift star formation and the local 
universe has recently been used to understand the observed population 
of faint dwarf galaxies in the Milky Way (MW) halo, we explore how well 
these nearby objects can probe the epoch of first light.
We construct a detailed, physically motivated model for the MW satellites 
based on the state-of-the-art {\it Via Lactea II} dark-matter simulations.  
Our model incorporates molecular hydrogen (H$_2$) cooling in low-mass 
systems and inhomogeneous photo-heating feedback during the internal 
reionization of our own galaxy.  We find that the existence of MW satellites
fainter than $M_{\rm V}\approx -5$ is strong evidence for H$_2$ cooling in 
low-mass halos, while satellites with $-5>M_{\rm V}>-9$ were
affected by hydrogen cooling and photoheating feedback.  The age 
of stars in very low-luminosity systems and the minimum luminosity of these 
satellites are key predictions of our model.  Most of the
stars populating the brightest MW satellites could have formed after
the epoch of reionization.  Our models also predict a significantly
larger dispersion in $M_{300}$ values than observed and a number of
luminous satellites with $M_{300}$ as low as $10^{6}\,\msun$.

\end{abstract}


\section{Introduction}

According to the standard cold dark matter (CDM) paradigm of cosmic
structure formation, massive objects such as the halo of our own Milky
Way (MW) grow hierarchically with smaller subunits collapsing earlier
and merging to form larger and larger systems over time.  Computer
simulations have long shown that this merging is incomplete and that
the dense cores of such progenitors may survive today as
gravitationally bound ``subhalos" within their hosts
\citep[e.g.][]{Moore99}.  Recently, state-of-the-art simulations have
revealed that present-day galaxy halos are extremely lumpy, filled
with tens of thousands of surviving substructures on all resolved mass
scales \citep{Diemand07,Diemand08,Springel08}.

The predicted subhalo counts vastly exceed the number of known dwarf
satellites of the MW, creating a ``missing satellite problem" whose
solution within the $\Lambda$CDM framework may lie both in the
luminosity bias that affects the observed satellite luminosity
function \citep{Koposov08,Tollerud08} as well as in the reduced star
forming efficiency predicted for small-mass, dwarf-sized substructure.
It is widely accepted that cosmic reionization may offer a plausible
mechanism for effectively inhibiting star formation in halos that are
not sufficiently massive to accrete warm intergalactic gas, and a
number of studies have attempted to interpret the observed population
of MW satellites using the process of early, external UV background
feedback
\citep{Bullock00,Benson02,Somerville02,Kravtsov04,Moore06,Strigari07,SG07,Madau08a,Busha09,Maccio09b}.

Lately, it has been suggested that a detailed quantitative resolution
of the ``missing satellite problem" may require some
``pre-reionization" suppression mechanism to avoid producing too many
faint Galactic dwarf spheroidals (dSphs).  Using results from the {\it
Via Lactea II} (VLII) simulation, \citet{Madau08b} showed that
thousands of surviving subhalos in the halo of the MW today have
progenitors massive enough for their gas to cool via excitation of
H$_2$ and fragment {\it prior to the reionization epoch}, which they
assumed occurred around $z\sim10$.  In addition, they found that star
formation in these objects must have been extremely inefficient
converting only a small fraction of their gas into stars or having
a top-heavy initial mass function (e.g. \citealt{Abel02}).  Similar
conclusions have been reached by \citet{Koposov09}.
 
Inspired by these results, we develop a detailed, astrophysically
motivated model for the formation of dSphs.  We consider a general
scenario in which $H_2$ fragmentation can shut off before the
suppression of atomic hydrogen cooling during reionization and include
post-reionization star formation in the largest subhalos.  This work
is unique in that it considers the possibility that the MW was
self-reionized from the inside out, which further suppresses the
amount of stars produced in progenitors with $M>10^8\,\msun$.  This
assumption is in agreement with observations that the mean-free-path
of ionizing radiation through intervening Lyman-limit systems (LLSs)
is much less, at $z>7$, than the $20 \rm Mpc$ distance between the
Virgo Cluster and the MW \citep{CAFG08}.
Using this physical model with the most recent observations of the
ultra-faint MW satellites found in the {\it Sloan Digital Sky Survey}
Data Release 5 (SDSS DR5) as a probe of both reionization and early
star formation physics, differentiates our work from previous studies
that focused on the properties of dSphs.  In this work we adopt
subhalo catalogs from the one billion particle VLII simulation, which
allow us to track the progenitors of surviving present-day MW
substructure far up in the merger hierarchy than done before.  The
unprecedented mass resolution, combined with the fossil signatures of
the reionization epoch in the Galactic halo, allows us to study gas
cooling in the early universe, star formation in the first generation
of galaxies, and the baryonic building blocks of today's galaxies.
  
This {\it{Paper}} is organized as follows.  In \S\ref{sec:model}, we
briefly describe the VLII simulation and develop our model.  In
\S\ref{sec:obs}, we compare the results with observations, examine
which observables best constrain model components, and determine model
parameters.  Finally, we summarize our conclusions and discuss how
they contribute to our understanding of reionization and high-redshift
star formation in \S\ref{sec:conc}.

\section{Basic Model}\label{sec:model}

The recently completeted VLII simulation \citep{Diemand08} uses just
over a billion dissipationless particles 
each weighting $4,100\,\msun$ to simulate
the formation of a $M_{200}=1.9\times10^{12}\,\msun$ MW-sized halo in
a {\it Wilkinson Microwave Anisotropy Probe (WMAP)} 3-year cosmology
\citep{Spergel07}.  It resolves $50,000$ subhalos today within the
host's $r_{200}=402$ kpc (the radius enclosing an average density
$200$ times the mean matter value) and tracks the merger history of
each subhalo that survives to the present epoch through $400$ time
slices between $z=27.54$ and today.  In this study, we analyze
catalogs that, for each surviving subhalo, give the number of
progenitors and their masses at a given redshift.  The details of the
progenitor determinations are given in \citet{Madau08b}.  Progenitors
are resolved down to masses of $10^5\,\msun$, almost a factor of $200$
better than the dissipationless simulation of the MW used by
\citet{GK06}.  This extra resolution allows us to incorporate physical
models involving H$_2$ cooling in very low-mass halos at extremely
high redshift.

Our model for assigning stars and luminosities to MW subhalos from the
VLII catalogs involves carefully charting the state of cosmic hydrogen
and a comparison of the cooling mass with the Jeans mass throughout
cosmic time.  We identify four epochs of star formation that can
contribute to the stellar populations of MW dSphs: (1) stars can form
at $z\sim 20$, well before reionization, in systems large enough for 
the cooling of molecular hydrogen; (2) \HI\ cooling produces stars before 
reionization that are responsible for photoheating the IGM;
(3) further cooling via \HI\ from reionization to $z\sim2$ produces
stars in subhalos large enough to hold onto their gas; and (4) stars
form in the last $10\,$Gyr from metal cooling.  In the following
subsections, we describe the process of star formation in each of the
above epochs following \citet{BL01}, and detail how we attach stellar
populations to the simulation subhalos.  In incorporating this model,
we use the set of best-fit cosmological parameters from the WMAP
5-year data release \citep{Komatsu09}, which are fully consistent with the previous results from WMAP3.

\subsection{Epoch of Molecular Hydrogen Cooling}\label{sec:H_2}

In the pristine early universe beyond $z\sim20$, cooling via H$_2$ is
efficient in systems with total halo masses as low as $M_{\rm
H_2}\sim5\times10^4\,\msun$, while the Jeans mass is lower by an order
of magnitude \citep{BL01}.  
The resulting first stars create a background of Lyman-Werner photons,
which both dissociates H$_2$ and acts as positive feedback to
replenish it \citep{RGS02a,RGS02b}.  Supernova explosions from these stars also
begins to enrich the IGM with the small amount of heavy metals
necessary for metal cooling.

Ultra-faint MW dwarf satellites are a natural and unavoidable
consequence of H$_2$ cooling in the pre-reionization universe
\citep{RG05,BR08}, and this mechanism can be responsible for the low
observed metallicities of these objects \citep{SF08}.  We take
advantage of the high mass resolution of VLII to trace the
pre-reionization progenitors of these ultra-faint dwarfs, for the
first time, down to an H$_2$ cooling mass as small as $10^5\,\msun$ to
make quantitative predictions about their abundance and observable
properties.

Motivated by \citet{Madau08b}, where an extremely small star formation
efficiency was invoked in $<10^7\,\msun$ objects at $z=11$ to avoid
overpopulating the faint end of the MW satellite luminosity function
(LF), we allow for the possibility that H$_2$ cooling is suppressed
earlier than \HI\ cooling \citep{HRL97,HAR00}.  Earlier suppression could explain the very
small efficiency of these small objects in a very natural way since
the halos able to cool via molecular hydrogen would be less numerous
at early times.  \citet{Whalen08} have suggested that this suppression
may occur around $z\sim20$ and result from supernova feedback.
  
We assume a redshift, $z_{\rm H_2}$, after which H$_2$ cooling is
quenched, and a mass threshold for halos, $M_{\rm H_2}$, below which
such cooling is inefficient.  We consider the VLII redshift snapshots
at $z=17.9$ and $z=23.1$ as possible values for $z_{\rm H_2}$ and
extract from the halo catalogs all surviving MW subhalos at $z=0$ with
progenitors above $M_{\rm H_2}$ at $z_{\rm H_2}$.  We further assume
that a fraction $f_b=\Omega_b/\Omega_M\approx0.16$ of the total mass
in each progenitor is in gas.  Due to the metal-poor state of the primordial gas,
the first generation of Population III stars has a top-heavy initial mass 
function (IMF) and does not survive to the present day.  Rather, the deaths of these 
stars seed the gas with traces of metals and spawn a new, metal-cooled 
stellar population with a Salpeter IMF.  While these metal-cooled stars are 
the ones with local relics seen today, the system initially must have satisfied the conditions for H$_2$ cooling.  We assume that, through this process, a fraction 
$f_{\rm H_2}$ of the initial gas is converted into stars with a Salpeter IMF; this 
factor accounts for any gas that may have been expelled by supernovae.
The low-mass stellar population will survive to the present day with a
visual luminosity of $M_{\rm V}=6.7$\footnote{We set a minimum stellar mass of $0.1\,\msun$ for the Salpeter IMF.  However, in principle, the cosmic microwave background creates a temperature floor to any cooling process that may in turn determine the minimum mass of stars at high redshifts \citep{BL03,Bromm04}.  In our discussion, we implicitly assume that stars at this lower limit would survive to the present day (or equivalently that this limit is below
$\sim 1 M_\odot$).  In this regime, the luminosity per unit mass is not
very sensitive to the value of the low-mass cutoff of the Salpeter mass
function.} per solar mass of initial stars
\citep{BC03,Madau08b}.  For each surviving subhalo, we sum up the
stars contributed by each progenitor at $z_{\rm H_2}$.  We assume that
these stars are quickly incorporated into the center of the forming
subhalo so that they are not stripped during subsequent mergers or
tidal encounters with the MW.

\subsection{Atomic Hydrogen Cooling Before Reionization}\label{sec:HI}

Next, we discuss the formation of stars in subhalos from
\HI\ cooling prior to the photoheating of the IGM at reionization.
This cooling process is very efficient only in gas with temperatures
above $T_4\equiv10^4\,K$.  Before the IGM photoheats to approximately
$T_4$, this temperature requirement limits \HI\ cooling to halos whose
virial temperatures exceed $T_4$ or, equivalently, to those with halo
masses above $M_4=10^8\,\msun\,\left[(1+z)/10\right]^{-3/2}$
\citep{BL01}.  Stars formed in this way are responsible for producing
the $N_{\rm clump}$ photons required on average to reionize each atom
of \HI\ in the Universe.  This happens when the collapse fraction of
halos with $M>M_4$ exceeds a critical threshold \citep{BL04}.  Thus,
the condition for the average reionization redshift of the universe,
$\bar{z}_{\rm rei}$, is:
\begin{equation}\label{eq:Ngamma}
N_{\rm clump}\,\Omega_b\,\rho_{\rm crit}=N_{\gamma}\,f_{\rm \HI}\,f_b\,\int_{M_4}^{\infinity}\,dM'\,M'\,\frac{dn}{dM'}(M',\bar{z}_{\rm rei}),
\end{equation}
where $dn/dM$ is the comoving mass function of halos, $N_{\gamma}$ is the number of ionizing photons produced per baryon in stars, $f_{\rm \HI}$ is the fraction of gas turned into stars by \HI\ cooling, and $\rho_{\rm crit}$ is the critical density of the universe today.  Here, $\Omega_b\,\rho_{crit}$ represents the total comoving mass density of baryons in the Universe to be reionized, not just those in underdense or mean-density regions.  Therefore, the standard factor $f_{\rm esc}$, the fraction of ionizing photons that escape into the IGM, does not appear on the right-hand side of equation~\ref{eq:Ngamma}.  Using the \citet{ST99} mass function for halos, we can solve for the combination $Q \equiv N_{\gamma}\,f_{\rm \HI}/N_{\rm clump}$ given any value of $\bar{z}_{\rm rei}$.  For $\bar{z}_{\rm rei}=11$, approximately the best fit WMAP5 value of the mean reionization redshift, we find $Q\sim100$.

After reionization, the IGM is rapidly photoheated to $\sim10^4\,$K.
Halos with masses less than the filtering mass, $M_F$, can no longer
hold onto their gas or accrete new baryonic material \citep{Gnedin00}.
While \citet{Koposov09} adopted the expression of \citet{Gnedin00} for
the remaining gas available for star formation, \citet{Busha09}
correctly pointed out that only cold gas can actually form stars.  The
situation is made worse by infalling gas whose temperature increases
by an extra order of magnitude during collapse.  Thus, after
reionization, only gaseous halos with virial temperatures above
$T_5=10^5\,K$ may cool via \HI\ to form additional stars (see
\S\ref{sec:after}).

As in \S\ref{sec:H_2}, we compile a list of surviving MW subhalos at
$z=0$ and all of their progenitors with $M>M_4$ at $\bar{z}_{\rm rei}$
and assign to each a stellar mass $M_{\star}$ and a luminosity (today,
after more than 13 Gyr of cosmic time) of $M_{\rm V}=6.7$ per solar mass of
stars distributed with a Salpeter IMF.  Our choice of IMF constrains
$N_{\gamma}=4000$ \citep{Bromm01}.  
As before, we assume that partial tidal stripping
of their hosts at later times does not remove these deeply embedded
stars.  Present-day subhalos with $M>M_4$ progenitors at $\bar{z}_{\rm
rei}$ as well as earlier progenitors at $z_{\rm H_2}$ with $M>M_{\rm
H_2}$, as outlined in \S\ref{sec:H_2}, have contributions to their
stellar populations from both epochs.

We now consider two alternatives for the reionization history of the
MW galaxy-forming region (MWgfr) that affect the calculation of
$M_{\star}$.  \citet{Alvarez09} suggest that
the MW was most likely ionized externally by radiation from the Virgo
Cluster.  However, they do not include the effect of intervening LLSs,
which absorb external UV photons and may allow the MWgfr to reionize
from the inside out instead.
Here we consider and present results for both scenarios.

If the MWgfr was reionized externally by the Virgo Cluster, we would
expect the ionization front to cross the region very quickly.  In this
case, we can assume that the region reionized promptly at
$\bar{z}_{\rm rei}$, and consider the subhalo progenitors at that
redshift.  The mass in stars formed by each subhalo progenitor at
$\bar{z}_{\rm rei}$ is the same as that assumed by \citet{Madau08b},
\begin{equation}\label{eq:MsEx}
M_{\star}=f_{\rm \HI, ex}\,f_b\,M_{\rm halo},
\end{equation}
where the subscript ``ex" denotes the assumption of external
reionization.  Considering only stars that formed through the \HI\
cooling of gas before $z=11$ and assuming a Salpeter IMF,
\citet{Madau08b} found that a star formation efficiency of $f_{\rm
\HI}\approx0.02$ for progenitors with $M_{\rm
halo}>7\times10^7\,\msun$ would reproduce the observed satellite LF.

On the other hand, if the MWgfr was ionized internally, different
parts of it would have reached the critical collapse fraction of
\HI-cooling halos necessary for reionization at different times due to
their respective overdensities \citep{BL04}.  Thus, there should be
fluctuations in the redshift of reionization within the MW itself
about the mean at $\bar{z}_{\rm rei}$, an effect that has not been
considered in the literature.  Each subhalo progenitor will
photoionize and reheat its own gas content at the time when it has
produced $N_{\rm clump}$ ionizing photons per hydrogen atom.  This
condition is met when
\begin{equation}\label{eq:MsIn}
M_{\star}=\frac{N_{\rm clump}\,g(\delta)}{N_{\gamma}\,\left(1-f_{\rm esc}\right)}\,f_b\,M_{\rm halo}=\frac{f_{\rm \HI}\,g(\delta)}{Q\,\left(1-f_{\rm esc}\right)}\,f_b\,M_{\rm halo},
\end{equation}
where $g(\delta)$ gives the ratio of $N_{\rm clump}$ in the overdense progenitor to the average $N_{\rm clump}$ in the universe.  The factor $\left(1-f_{esc}\right)$ selects only those photons that do not escape from the overdense progenitor and can be used to photoionize it.  Equation~\ref{eq:MsIn} tells us how much stellar mass to assign aprogenitor of mass $M_{\rm halo}$.  Each progenitor, A, has its own $M>M_4$ progenitors, B, that form stars at redshift $z'_{\rm rei}>\bar{z}_{\rm rei}$, before system A collapses at $\bar{z}_{\rm rei}$.  Baryonic material from the region that formed the A progenitor not converted into stars at $z'_{\rm rei}$ is photoionized before it can be incorporated into A at $\bar{z}_{\rm rei}$ and cannot form stars unless progenitor A eventually reaches a mass of $M_5$, the mass corresponding to a virial temperature of $T_5$ or a circular velocity of $V_5=52\,\kms$.  However, we expect most of the gas in B progenitors with $M>M_4$ at $z'_{\rm rei}$ to form stars.

Accounting for the inside-out morphology of the MWgfr reionization provides a natural explanation for the low values of $f_{\rm \HI, ex}$ found by several studies \citep{Madau08b,Koposov09,Busha09}, which assumed a single value for the reionization redshift of the entire MWgfr.  Since stellar mass is a linear function of halo mass in both equations~\ref{eq:MsEx} and~\ref{eq:MsIn}, both morphologies fit the same observed LF  when
\begin{equation}\label{eq:fHIex}
f_{\rm \HI, ex}=\frac{f_{\rm \HI}\,g(\delta)}{Q\,\left(1-f_{\rm esc}\right)}.
\end{equation}
If $f_{\rm \HI}\,g(\delta)/\left(1-f_{\rm esc}\right)$ is of order
unity at $\bar{z}_{\rm rei}=11$, we find that $f_{\rm \HI,
ex}\approx0.01$.  We can understand this physically by considering
how, at the redshift of star formation suppression, a lower efficiency
is equivalent to a smaller amount of matter in collapsed objects
(smaller $M_{\rm halo}$ in Eq.~\ref{eq:MsEx}) in terms of calculating
how much stellar mass is produced.  We see from this that, for a given
average redshift of reionization in the MWgfr, the inside-out and outside-in 
morphologies fit the luminosity function equally well but result in different 
interpretations of the fitting parameters.

\subsection{Star Formation After Reionization}\label{sec:after}

We assume that cosmic reionization and the photoheating of the MWgfr
completes rapidly after $\bar{z}_{\rm rei}$.  Subsequently, only halos
that have accumulated a mass of $M_5>1$ are able to hold onto their gas
long enough for it to form further stars via \HI-cooling.  We assume
that this star formation is quenched when the subhalo begins to
interact with the MW host.  The merger histories of each surviving MW
subhalo at $z=0$ show only eight subhalos that achieved a maximum
circular velocity greater than $V_5$ at some point in their histories.
We determine the redshift at which the maximum circular velocity
reaches its peak.  The subsequent reduction is due to 
tidal interactions during infall into the MW host with both neighboring 
dwarfs and the host itself \citep{Kravtsov04} that we assume coincide 
with the quenching of star formation through ram pressure.  In
agreement with \citet{Koposov09}, we find that this mass loss
typically occurs around $z=2-4$ (in five of our eight subhalos) but can
happen as early as $z=8$ and as late as $z=1.6$.

We add an additional mass of $f_5\,f_b\,M_{\rm max}$ worth of stars
minus the mass of any stars formed in either of the first two epochs
to each subhalo whose peak circular velocity exceeds $V_5$.  Here,
$f_5$ is an efficiency parameter for star formation during this epoch
that takes into account how much of the hot gas is cooled to form
stars and how much of the gas mass is removed during infall.
$M_{max}$ is the virial mass corresponding to the maximum value of the
circular velocity for each subhalo.  We calculate the age of these new
stars from the redshift at which the subhalo attains its maximum
circular velocity and use this age to calculate the luminosity of each
solar mass of stars from \citet{BC03}.
\subsection{Recent Star Formation}\label{sec:metals}

The final episode of star formation that we include in our model occurs in the last $10\,\rm {Gyr}$ since $z=1.6$.  Since we are interested primarily in the physics of the early universe, we do not attempt to develop here a detailed model for the production of these stars but presume that metal cooling is involved and use the observations of \citet{Orban08} to add additional young stars in a stochastic way.

For each of the classical, pre-SDSS MW satellite, \citet{Orban08} have measured $f_{10G}$, the fraction of stellar mass produced in the last $10\,\rm{Gyr}$, and $\tau$, the mass-averaged stellar age.  They find that the metallicities of the ultra-faint, SDSS satellites are so low that no star formation is expected for these in the last $10\,\rm{Gyr}$.  As we will show, subhalos that achieve a maximum circular velocity of at least $V_5$ are associated only with the classical MW satellites studied by \citet{Orban08}, while stars that formed prior to reionization populate both the classical and SDSS dSphs.  Therefore, we set $f_{10G}=0$ for subhalo progenitors that only formed stars before reionization to be consistent with these metallicity observations.  However, setting this constraint only for progenitors with H$_2$-induced cooling does not significantly change our results.

The cumulative probability distribution of $f_{10G}$ is approximately linear with a linear fit reaching unity at $f_{10G}=0.89$ \citep{Orban08}.  Thus, for each subhalo with stars that formed after reionization, as outlined above, we select a value for $f_{10G}$ at random from a flat probability distribution within the interval $\left[0.00,0.89\right]$.  The mass of stars produced since $z=1.6$ is, therefore, given by $f_{10G}/(1-f_{10G})$ times the mass of stars produced from the mechanisms outlined in the previous three subsections.

If we pretend that all of the stars formed in the MW satellites prior
to $z=1.6$ have an age of $14\,\rm{Gyr}$, we can use the measured
values of $f_{10G}$ and $\tau$ to estimate the mass-weighted average
age of those stars formed in the last $10\,\rm{Gyr}$.  We find that
the stars that formed after $z=1.6$ in each satellite have ages
between $1\,\rm{Gyr}$ and $8\,\rm{Gyr}$ with an average around
$5\,\rm{Gyr}$.  Since the luminosity per solar mass of a stellar
population does not vary much in this age range \citep{BC03}, we assume
a fixed age of $5\,\rm{Gyr}$ for the recent stars formed in each of
our subhalos.  \citet{BC03} give the luminosity of a stellar
population with this age as $M_{\rm V}=5.8$ per solar mass.

\section{Comparison with Observations}\label{sec:obs}

We are now at a position to compare the properties of the luminous subhalos in
our model with observations of MW dwarf satellites.  These include
values for Segue II, the newest MW satellite discovered in the SDSS
data \citep{Belokurov09}, but exclude Leo T which, at a distance of
$417\,\rm{kpc}$ from the Sun, is outside the viral radius of the VLII
simulation.  The simulation abundance calculations represent mean
values about which there is Poisson scatter.  However, for clarity in
the plots, we instead assign Poisson errors on the observational data
as if the observed count were the mean value.  We ignore the
variations that would expected in different cosmological realizations
of the MW and leave this analysis to future work.

In \S\ref{sec:completeness}, we first outline the observational biases of our observed sample of satellites and describe how we apply a similar bias to the simulated population to account for the completeness of the sample.  We then proceed to compare the simulated and observed LFs of satellites in \S\ref{sec:Lfunc}, and investigate how this important observable constrains our model.  Finally, we consider how well our model reproduces other observables in \S\ref{sec:otherobs} such as the radial distribution of satellites within the MW halo, the mass of a satellite within $300\,\rm{pc}$ of its center, and a satellite's mass-to-light ratio.

\subsection{Observational Completeness}\label{sec:completeness}

Our sample includes two sets of MW satellites: those discovered prior to SDSS and the ultra-faint sample found in the SDSS DR5.  However, the observational biases for the two samples are not the same.  Not only is the SDSS sample complete only to a given surface brightness threshold, but the SDSS footprint only covers $20\%$ of the sky.  Rather than apply corrections to the observed distribution functions of satellites to account for the total population as in \citet{Tollerud08}, we follow \citet{Madau08b} in adjusting our simulated population to represent those objects that would be detected by SDSS.

The observational completeness of the SDSS DR5 was modeled by
\citet{Koposov08} and \citet{Tollerud08} by defining the maximum
radius at which a satellite of a given visual magnitude would be
observed.  This radial threshold is given by:
\begin{equation} \label{eq:rmax}
r_{\rm max}=\left({\frac{3}{4\pi f_{\rm DR5}}}\right)^{1/3}\,10^{(-0.6\,M_{\rm V}-5.23)/3}\,{\rm Mpc}, 
\end{equation}
where $f_{\rm DR5}=0.194$ is the fraction of the sky covered by DR5.  We exclude each subhalo in our simulation with a distance beyond this threshold for its particular luminosity.  Moreover, when plotting the luminosity and radial distribution functions of satellites in \S\ref{sec:Lfunc} and \S\ref{sec:otherobs}, we correct these distribution by an additional factor of $f_{\rm DR5}$.  The observed distributions are also corrected so that each classical MW satellite contributes only $f_{\rm DR5}$ to the total abundance, while each SDSS satellite contributes fully.  However, all observed satellites contribute equally when calculating the fractional Poisson errors in the distribution functions.

\subsection{The Luminosity Function}\label{sec:Lfunc}

First, we would like to compare our simulated LF of MW satellites to the
observed distribution and explore the importance of this observable
for constraining model parameters.

\begin{figure}
\begin{center}
\includegraphics[width=\columnwidth]{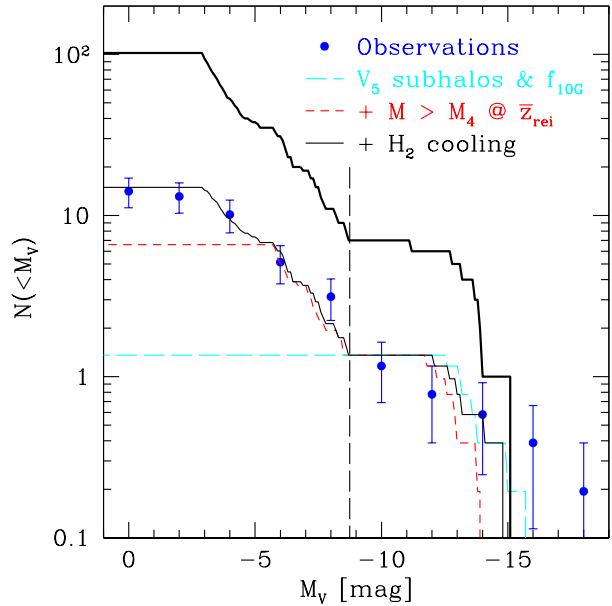}
\caption{\label{fig:Lfunc} The luminosity function (LF) of MW satellites.  
Thin lines show abundances in the SDSS footprint with the DR5 selection threshold.  The observed function is shown by the blue points with the non-SDSS
objects each contributing a fractional amount $f_{\rm DR5}=0.194$ to
the total.  The curves show the theoretical predictions including
successively more elaborate models.  The model given by the
long-dashed, cyan curve illuminates only those VLII subhalos with
maximum circular velocities that exceed $V_5$ at some point in their
histories and continue to form stars after reionization via atomic
hydrogen cooling (with an efficiency of $f_5=0.02$) and metal cooling
in the last $10\,{\rm Gyr}$ ($f_{10G}\neq0$).  The short-dashed, red
curve includes star formation in subhalo progenitors more massive than
$M_4$ at $\bar{z}_{\rm rei}=11.2$ assuming a single redshift of
reionization for the entire MWgfr and $f_{\rm \HI, ex}=0.02$.  The thin 
solid, black curve fitting the faint end of the observed LF
additionally takes into account molecular hydrogen cooling, prior to
suppression at $z_{\rm H_2}=23.1$, in progenitors more massive than
$M_{\rm H_2}=10^5\,\msun$ with $f_{\rm H_2}=0.4$.  The thick, solid, black curve represents the same model as the thin, solid one, but now all satellites in the MW halo out to the virial radius are show, not just those that meet the SDSS criteria.  The long-dashed vertical line demarcates the luminosities at which pre- vs. 
post-SDSS satellites are observed.  }
\end{center}
\end{figure}

The observed and simulated LFs are shown in Figure \ref{fig:Lfunc}.  We present multiple theoretical curves, each of which includes additional layers of the model to demonstrate the relationship between the model and the predictions.  For example, taking into account only subhalos that reach a maximum circular velocity of at least $V_5$ at some point in their histories (see \S\ref{sec:after} and \S\ref{sec:metals}) with $f_5=0.02$ roughly reproduces the LF of pre-SDSS satellites.  Recent, high-metallicity star formation in the last $10\,{\rm{Gyr}}$ introduces some stochasticity in the theoretical LF, but the fluctuations are smaller than the observed errors.  However, this limited model fails to account for observed satellites dimmer than $M_{\rm V}\approx-9$.  The addition of star formation in progenitors with $M>M_4$ before reionization at $\bar{z}_{\rm rei}=11.2$ extends the agreement between the theoretical and observed LFs down to $M_{\rm V}\approx-6$ when we assume $f_{\rm \HI, ex}=0.02$ in an external reionization model or $N_{\rm clump}\,g(\delta)\,\left(1-f_{\rm esc}\right)\approx30$ in an inside-out model.  Additionally, the inclusion of pre-reionization stars lowers the post-reionization efficiency to $f_5=0.01$.

However, it is only with the inclusion of stars in low-mass molecular hydrogen cooling systems that the model output reproduces observations of ultra-faint satellites.  This element of our model and its ability to explain and be constrained by the data is a major new development in this work that distinguishes it from other studies in the literature.  In Figure \ref{fig:Lfunc}, we assumed $z_{\rm H_2}=23.1$ and $M_{\rm H_2}=10^5\,\msun$ with $f_{\rm H_2}=0.4$.  The model not only gives the correct abundance of satellites brighter than the faintest known object, Segue 1, it also provides an explanation for why no fainter objects have been found.  To produce stars, our model requires a subhalo to have at least one progenitor with mass above the cooling mass for molecular hydrogen.  The fewest stars are produced when the subhalo has exactly one of these progenitors that is just above $M_{\rm H_2}$.  In this case, the mass in stars is $M_{\star}=f_{\rm H_2}\,f_b\,M_{\rm H_2}\approx6600\,\msun$, and the luminosity today is $M_{\rm V}>-2.8$.  The existence of a satellite as dim as Segue 1 is within the uncertainty of our approximations.  While a precise measurement of the lower limit on the luminosity of MW satellites in the future will provide a tighter constraint on early star formation, our model implies that the exact value cannot be significantly below what has already been observed.  

\begin{figure}
\begin{center}
\includegraphics[width=\columnwidth]{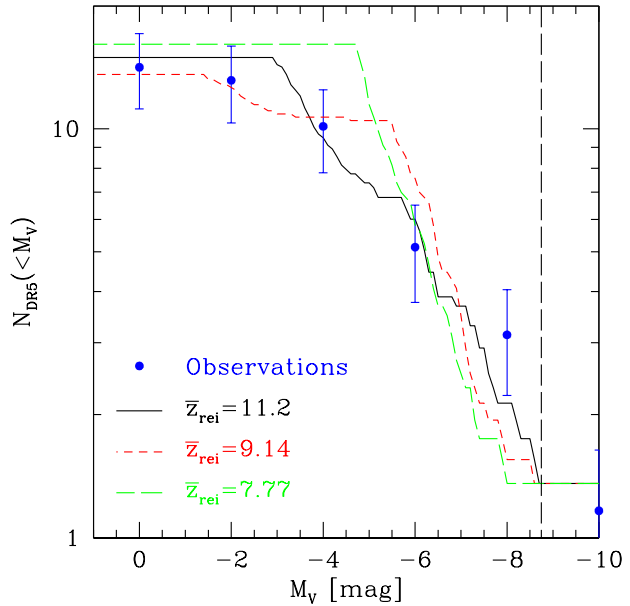}
\caption{\label{fig:Lfunc1} The faint end of the luminosity function
of MW satellites in the SDSS footprint with the DR5 selection
threshold.  The observed function is shown by the blue points with the
non-SDSS objects each contributing a fractional amount $f_{\rm
DR5}=0.194$ to the total.  The curves show the theoretical predictions
for different values of $\bar{z}_{\rm rei}$ with $z_{\rm H_2}=23.1$
and other model parameters optimized to produce the best agreement
with luminosity function data.  The solid, black; short-dashed, red;
and long-dashed, green curves have $\bar{z}_{\rm rei}=11.2$, $9.14$,
and $7.77$, respectively.  The long-dashed vertical line demarcates
the luminosities at which pre- vs. post-SDSS satellites are observed.
}
\end{center}
\end{figure}

We emphasize this intriguing result that star formation from different time periods, before the suppression of H$_2$ cooling and before and after reionization, are required to fit the theoretical LF to the data at the faint, middle, and bright ends, respectively.  Although some objects do have stars from multiple epochs, they are typically dominated by the processes in a single part of the model.  This helps prevent degeneracy among all of our different parameters and allows us to learn about star formation parameters in each stage almost independently of the others.

So far, we have produced results by fixing $z_{\rm H_2}=23.1$ and $\bar{z}_{\rm rei}=11.2$ and varying the efficiencies and mass thresholds to obtain agreement.  However, by allowing for different values of the suppression redshifts, we can arrive at new sets of parameters that allow the model to fit the observed LF.  

Figure \ref{fig:Lfunc1} compares the simulated LFs with reionization fixed at three consecutive VLII slices for which progenitor analysis is available: $\bar{z}_{\rm rei}=7.77$, $9.14$, and $11.2$.  For each of these redshifts, we find $Q=28$, $46$, and $130$.  We have fixed $z_{\rm H_2}=23.1$ and $M_{\rm H_2}=10^5\,\msun$ but varied $N_{\rm clump}=N_{\gamma}\,f_{\rm \HI}/Q$ and $f_{\rm H_2}$ to find the values resulting in the closest fit to the data.  For convenience we define $N'_{\rm clump}\equiv N_{\rm clump}\,g(\delta)/\left(1-f_{\rm esc}\right)$ and $f'_{\rm \HI} \equiv f_{\rm \HI}\,g(\delta)/\left(1-f_{\rm esc}\right)$.  Our sets of fit parameters, then, are $\left(\bar{z}_{\rm rei},N'_{\rm clump},f'_{\rm \HI},f_{\rm H_2}\right)=\left\{(7.77,7,0.05,<0.001),(9.14,17,0.2,0.1),(11.2,30,1,0.4)\right\}$; the values of $f_{\rm \HI, ex}$ for an external reionization scenario can be calculated from equation~\ref{eq:fHIex}.

At earlier times, less mass in the Universe is in collapsed objects above the cooling mass.  This results in a monotonically increasing value of $Q$ with $\bar{z}_{\rm rei}$ since each baryon in these objects must ionize a larger number of hydrogen atoms.  This requires the universe to be {\it less} clumpy for fixed $f_{\rm \HI}$.  However, the reduction in the mass contained in each present-day subhalo's progenitors as $\bar{z}_{\rm rei}$ increases requires larger values of $f'_{\rm \HI}$ to produce the same observed stellar mass today.  This effect produces global reionization more efficiently and allows {\it increased} clumping.  Since $f'_{\rm \HI}$ increases faster with $\bar{z}_{\rm rei}$ than does $Q$, the net result is an increasing $N'_{\rm clump}$.

\begin{figure}
\begin{center}
\includegraphics[width=\columnwidth]{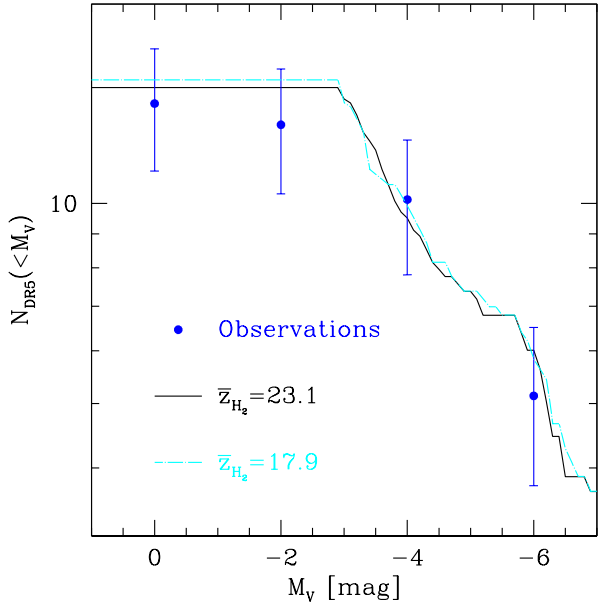}
\caption{\label{fig:Lfunc2} 
Same as Figure \ref{fig:Lfunc1} except we have fixed $\bar{z}_{\rm rei}=11.2$ and the solid, black and dot-dashed cyan curves show theoretical predictions for sets of parameters given $z_{\rm H_2}=23.1$ and $17.9$, respectively. 
}
\end{center}
\end{figure}

Because later reionization allows for star formation in more objects near the cooling threshold before \HI\ cooling is suppressed, star formation in molecular hydrogen cooling halos at earlier redshifts is allowed to be less efficient to produce roughly similar LFs.  This is responsible for the reduced values of $f_{\rm H_2}$ that we for lower $\bar{z}_{\rm rei}$.  The value $f_{\rm H_2}<0.001$ effectively means that the contribution of stars in molecular hydrogen cooling systems before $z_{\rm H_2}$ to the luminosity of MW satellites is negligible; this results in an absence of satellites with luminosities fainter than $M_{\rm V}\approx-5$.

While all three considered values of $\bar{z}_{\rm rei}$ produced similar totals for the number of observed MW satellites, the LF slope between $-5>M_{\rm V}>-9$ seems to favor earlier reionization around $\bar{z}_{\rm rei}=11$.

We also considered the effect of varying $z_{\rm H_2}$ and examined
two VLII slices at $z=17.9$ and $23.1$.  We fixed the reionization
parameters to be those best fit for $\bar{z}_{\rm rei}=11.2$, but
varied $M_{\rm H_2}$ and $f_{\rm H_2}$ to find the best agreement with
the LF data.  Figure \ref{fig:Lfunc2} shows the resulting LFs and
compares them with observations.  The model with parameters set at
$\left(z_{\rm H_2},M_{\rm H_2}/\msun,f_{\rm
H_2}\right)=(23.1,10^5,0.4)$ produces almost identical results as that
with $(17.9,10^6,0.04)$.  Allowing more time for stars to form before
the suppression of H$_2$ cooling is compensated for by a larger
cooling mass threshold and a smaller star formation efficiency.  The
minimum cooling masses for the redshifts we considered are around the
minimum masses typically assumed in simulations of feedback from the
first stars \citep{Whalen08}.

\subsection{Radial Distribution, $M_{300}$, and $M/L$}\label{sec:otherobs}

\begin{figure}
\begin{center}
\includegraphics[width=\columnwidth]{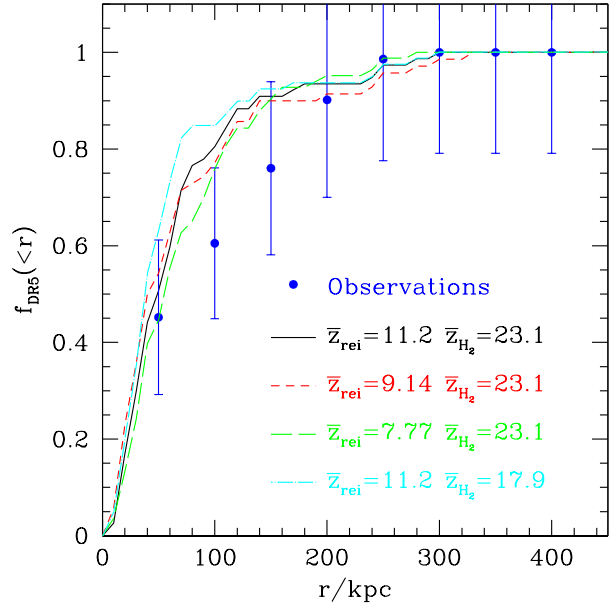}
\caption{\label{fig:Rfunc} 
The radial distribution function of MW satellites in the SDSS footprint with the DR5 selection threshold.  The profile is plotted as the fraction of the total within a given radius.  The observed profile is given by the blue points with the non-SDSS objects each contributing an amount $f_{\rm DR5}=0.194$ to the total.  The curves show the theoretical predictions for different pairs of fixed $(\bar{z}_{\rm rei}, z_{\rm H_2})$ with other model parameters optimized to produce the best agreement with luminosity function data.  The solid, black curve has $(11.2,23.1)$, the short-dashed, red curve $(9.14,23.1)$, the long-dashed, green curve $(7.77,23.1)$, and the dot-dashed, cyan curve $(11.2,17.9)$.
}
\end{center}
\end{figure}

Given the degeneracies of various degrees that we have seen in the LF
for some sets of model parameters, we attempt to use the other
observables to learn more about the early universe.  
Figure \ref{fig:Rfunc} shows the radial distribution throughout the MW
of the observed and simulated satellites for different pairs of fixed
$(\bar{z}_{\rm rei}, z_{\rm H_2})$ with other model parameters
optimized to produce the best agreement with luminosity function data.
We find that all of the $z_{\rm H_2}=23.1$ models are fairly
consistent with the radial data deviating non-negligibly only for
satellites in the range $50<r/{\rm kpc}<100$ of the MW center.  The
$(\bar{z}_{\rm rei}, z_{\rm H_2})=(11.2,17.9)$ prediction for the
radial distribution rises more sharply than expected for $r<100\,{\rm
kpc}$, but the statistical strength of this deviation is by no means
overwhelming.

We further explored whether these parameter sets were distinguishable in the distribution function of the maximum circular velocity of the subhalos.  This property is difficult to measure observationally from the velocity dispersion of real satellites at a given radius from the center of the object.  However, no significant difference was found among the distributions for model parameters that also fit the LF data.

In addition to considering distribution functions, we would also like to test for agreement between our model and observations of specific satellite properties.  The mass enclosed within $300\,{\rm pc}$, $M_{300}$, as a function of luminosity has been studied using both observational \citep{Mateo98,Gilmore07,Strigari08} and theoretical \citep{Li09,Koposov09,Maccio09a} approaches that hinted at a characteristic mass scale for MW satellites of about $10^7\,\msun$.  Both the relatively constant value of $M_{300}$ and the extreme observed mass-to-light ratios of the faintest satellites are properties of the population that we would like our model to reproduce, however, we keep in mind that both $M_{300}$ and the total mass from which mass-to-light ratios are calculated are difficult to determine observationally.

\begin{figure}
\begin{center}
\includegraphics[width=\columnwidth]{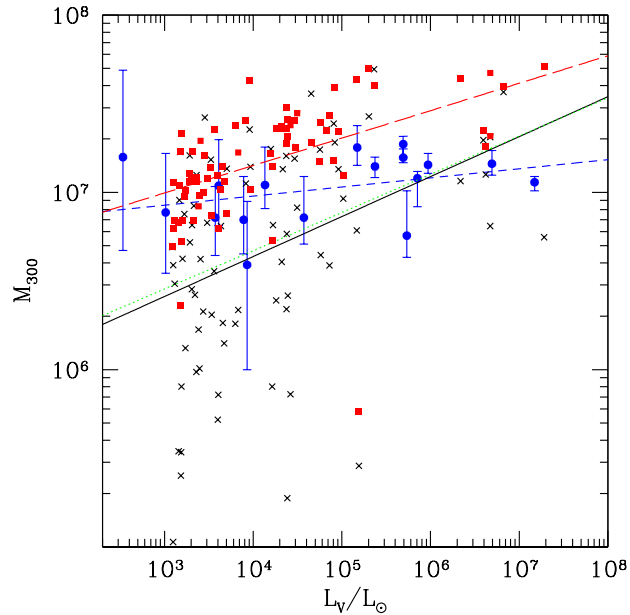}
\caption{\label{fig:M300} 
Scatter plot of $M_{300}$, the mass within $300\,{\rm pc}$ of the satellite center, versus satellite visual luminosity, $L_{\rm V}$.  Known MW satellites are represented by blue circles, while black x's denote illuminated subhalos from our $(\bar{z}_{\rm rei}, z_{\rm H_2})=(11.2,23.1)$ model with $M_{300}$ measured directly from the simulation.  Red squares show values of $M_{300}$ measured, for each subhalo, at the time when the evolution in its maximum circular velocity has reached its peak and assuming an NFW profile.  The short-dashed blue, solid black, dotted green, and long-dashed lines represent power-law fits of the $M_{300}-L_{\rm V}$ relation respectively from observations, directly from the simulation, assuming NFW profiles for simulated subhalos today, and assuming NFW profiles for the subhalos when their maximum circular velocities peak.
}
\end{center}
\end{figure}

\begin{figure}
\begin{center}
\includegraphics[width=\columnwidth]{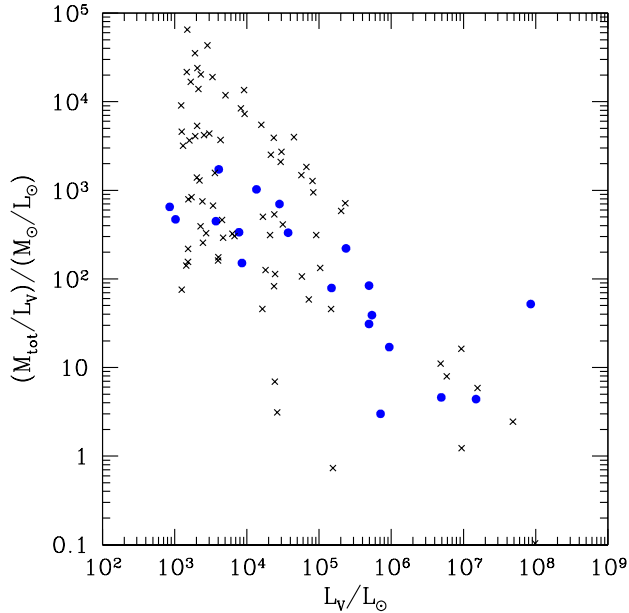}
\caption{\label{fig:MtoL} 
Scatter plot of the mass-to-light ratio of MW satellites versus their visual luminosities.  Known MW satellites, whose ratios are calculated based on the mass within stellar ``tidal" radii, are represented by blue circles.  Black x's denote illuminated subhalos from our $(\bar{z}_{\rm rei}, z_{\rm H_2})=(11.2,23.1)$ model whose mass-to-light ratios were calculated using the simulated mass within their dark matter tidal radii.
}
\end{center}
\end{figure}

In Figure \ref{fig:M300}, we present measurements of $M_{300}$ for MW satellites from \citet{Strigari08} and compare them with the output from our $(\bar{z}_{\rm rei}, z_{\rm H_2})=(11.2,23.1)$ model.  Figure \ref{fig:MtoL} similarly compares the observed mass-to-light ratios for each satellite \citep{Mateo98, SG07, Martin07} with the value calculated using the simulated mass within its tidal radius.  Both $M_{300}$ and the mass-to-light ratio are plotted as a function of satellite V-band luminosity taken from \citet{Tollerud08}, whose values are a little more up-to-date than those in \citet{Strigari08}.  While we continue to apply the SDSS DR5 selection criterion from equation~\ref{eq:rmax}, we plot all of the illuminated subhalos across the sky from our model in both figures rather than select only $20\,\%$ in the SDSS field-of-view.  However, we simply omit known satellites for which we were unable to find reliable estimates of the relevant data.

While our model predicts a fairly constant $M_{300}$ as a function of luminosity (a variation of about an order-of-magnitude over six decades in luminosity), there is still some disagreement at low luminosities where we find a lower average $M_{300}$ and greater scatter than observed.  We find that values of $M_{300}$ measured directly from the simulation match very well with those obtained assuming, for each satellite, an NFW profile \citep{NFW97} fit to its simulated maximum circular velocity and the radius at which this maximum is reached.  A power-law fit to the resulting scatter plot of $M_{300}$ vs. V-band luminosity, $L_{\rm V}$, of the form $M_{300} = \beta\,L_{\rm V}^\alpha$ yields $\alpha = 0.22$ for both the simulated and NFW values of $M_{300}$.  This is much steeper than the value of $\alpha = 0.05$ for the \citet{Strigari08} measured values of $M_{300}$ with the up-to-date values of $L_{\rm V}$.

Assuming that the NFW profile is a good fit for all subhalos at all
times, we also calculated $M_{300}$ for each subhalo at the redshift
when the evolution in its maximum circular velocity reaches its peak.
The difference between peak and present values should give us an idea
of how much tidal stripping changes the mass estimates.
\citet{Maccio09a} recently suggested that tidal stripping would result
in a flattening of the $M_{300}-L_{\rm V}$ relation.  However, our results
show a further steepening as more low-luminosity subhalos lose a
significant amount of their mass than do high luminosity ones.  We
find $\alpha = 0.15$ for the early values of $M_{300}$.  While it is
true that large, low-concentration systems lose more total mass to
tidal stripping than smaller ones do, they are actually denser and
more robust to stripping at a fixed radius.  Additionally, if the smaller subhalos are also initially fainter, they must be deeper in the inner halo to be detected (i.e. have smaller $r_{\rm max}$) where tidal effects are stronger, while larger, brighter halos can be detected even at large radii where tides are weak.

Of course, an NFW profile is not always a good fit for all subhalos at
all times.  For example, if the peak in the maximum circular velocity
is reached during a merger, one can get a rather large radius as well,
which results in a low concentration and low $M_{300}$. The actual
mass distribution during a merger is quite different from NFW, and the
true $M_{300}$ would be higher.  This explains why a few of the
subhalos plotted in Figure \ref{fig:M300} appear to gain mass from the
peak until today.  However, if we were able to plot the true, larger
values of $M_{300}$ at the time of the peak circular velocity for all
subhalos, it would only show a greater steepening between the epoch
when the peak is reached and today.

The tension at low luminosities between our predictions and the measurements of \citet{Strigari08} may well be explained by the difficulty in determining $M_{300}$ observationally.  The tracer stars in many small systems do not extend past $100\,{\rm {pc}}$ and converting the mass within this radius to that inside $300\,{\rm {pc}}$ requires assumptions about the dark matter distribution (e.g. shape, orientation, 
density profile) and about the orbits of the stars.  Additionally, interlopers and/or undifferentiated binary stars could systematically skew velocity dispersion
measurements, especially in those satellites where the dispersion is small 
($\sim 5\,{\rm km/s}$).  These errors could significantly inflate the mass of 
low-mass systems.  

While the observed mass-to-light ratios are calculated from the mass within the stellar ``tidal" radius, theoretical values from the simulation are computed using the mass within the dark matter tidal radius of the subhalo.  This means that the model predictions are upper limits on the observed values.  This is the best we can do without modeling the radial distribution of the stars within subhalos.  The two definitions of mass-to-light ratio agree only in systems where the stars and the dark matter are truncated by tides at the same radius.  Despite this difference, the model roughly reproduces the slope and amplitude of the relation between satellite mass-to-light ratio and visual luminosity.  We do appear to produce extra faint halos with ratios even larger than observed, which most likely is a result of the definitional difference between the observed and theoretical values.

These results show that our model not only produces the correct abundance of luminous subhalos at each luminosity, but it also gives subhalos with approximately the correct physical properties for their luminosity.  That is, the simulated subhalos that correspond to the brightest satellites or to the ultra-faint satellites in terms of their abundance have the same physical properties as the brightest or faintest satellites, respectively.  Our faint objects do not, for example, have the physical properties, such as $M_{300}$ or mass-to-light ratio, of the brightest satellites.  This consistency increases confidence in our model.

Unfortunately, the different sets of model parameters we have been considering produce the about same range of $M_{300}$ or mass-to-light ratio for a given luminosity.  Of course, certain luminosities are under- or over-populated by points in different models, but these differences are better represented as differences in the luminosity function as we have described in \S\ref{sec:Lfunc}.  Therefore, these scatter plots are less useful for learning about reionization and star formation in the early universe.

\section{Discussion and Conclusions}\label{sec:conc}

In this {\it{Paper}} we have shown that a physical model for star formation
applied to the subhalos of a high-resolution galactic N-body
simulation can, indeed, explain the full observed population of Milky
Way (MW) dwarf satellites.  
While \citet{Maccio09b} reached similar
conclusions, they were not able to reproduce the faint end of the
luminosity function (LF) because their model did not include molecular
hydrogen cooling in halos with masses below $10^8\,\msun$.  
Our inside-out model of the reionization of the MW galaxy forming region (MWgfr) is characterized by the star formation efficiency $f_{\rm H_2}$, cooling mass $M_{\rm H_2}$, and suppression redshift $z_{\rm H_2}$ of H$_2$ cooling, the normalized efficiency of \HI\-cooled stars $f'_{\rm \HI}$, normalized IGM clumping $N'_{\rm clump}$, and the mean redshift of reionization $\bar{z}_{\rm rei}$, and the star formation efficiency $f_5$ of massive systems capable of \HI\ cooling after reionization.  Here $N'_{\rm clump}
\equiv N_{\rm clump}\,g(\delta)\left(1-f_{\rm esc}\right)$ and
$f'_{\rm \HI} \equiv f_{\rm \HI}\,g(\delta)/\left(1-f_{\rm
esc}\right)$.
Our results are slightly at odds with recent measurements from the literature of $M_{300}$ for ultra-faint systems, but we anticipate that measurement uncertainties may have resulted in the difference (see \S\ref{sec:otherobs}).

In \S\ref{sec:Lfunc}, we used our model to explore what the observed LF of MW
satellites could tell us about the early universe 
and discovered that satellites of different luminosities
give clues about star formation at different epochs:

\begin{enumerate}

\item MW satellites with luminosities fainter than $M_{\rm
V}\approx-5$ can be explained by the inclusion of second-generation 
stars in low-mass halos above a cooling threshold of 
$M_{\rm H_2}\sim10^{5-6}\,\msun$ that were initially able to cool via molecular
hydrogen in the very early universe.

We took advantage of the VLII's high resolution to probe this process
in very small systems and allowed for the possibility that a mechanism
other than reionization was responsible for its suppression in the
early universe.  We have shown that our theoretical subhalos illuminated
in this way are very faint, metal-poor (due to their very early
creation), and have extreme mass-to-light ratios and the correct
abundance to be responsible for the faintest population of satellites
discovered in the SDSS DR5.

Using observations of these ultra-faint satellites to learn about the
star formation process, we have found that molecular hydrogen cooling
systems convert a fraction $f_{\rm H_2}$ of their initial gas into stars before 
star formation is suppressed very early in cosmic history at 
$z_{\rm H_2}$.  While the observed LF could not distinguish between 
models with $z_{\rm H_2}=23.1$ and $z_{\rm H_2}=17.9$, the radial 
distribution of these objects somewhat favors earlier suppression.  The best fit values of
the cooling mass and star formation efficiency for each value of
$z_{\rm H_2}$, assuming cosmic reionization at $\bar{z}_{\rm rei}=11.2$, are 
$\left(z_{\rm H_2},M_{\rm H_2}/\msun,f_{\rm H_2}\right)=\left\{(23.1,10^5,0.4),(17.9,10^6,0.04)\right\}$.  Only if $\bar{z}_{\rm rei}\la8$ is no H$_2$ cooling in low-mass 
systems required to reproduce the observed MW satellite LF.
Confidence in our models increases when we note that our values of the
cooling threshold are consistent with what has been already been
assumed in simulations of the formation of the first stars
\citep{Whalen08}.  Our results put further constrains on input
parameters for these studies.

The physical difference between the two sets of $\left(z_{\rm
H_2},M_{\rm H_2}/\msun,f_{\rm H_2}\right)$ values that we considered
may not be stark.  The fact that $f_5$ increases for an earlier
$z_{\rm H_2}$ may suggest a similarity with the inside-out
reionization model of \S\ref{sec:HI} where progenitors that reionized
before the rest of the MWgfr have a star
formation efficiency much great than in an in external reionization
scenario.  The similarities are intriguing, but we leave this problem
to future work.

For either value of $z_{\rm H_2}$ that we considered, ultra-faint satellites are clearly the sites of the very first and oldest in the universe, and we agree that searches for these stars should be targeted there \citep{Kirby08,Frebel09}.  While we do not expect stellar ages from future surveys precise enough to distinguish between formation at $z=20$ and $z=8$, the absence of any metal-poor stars older than $13\,{\rm Gyr}$ would represent a problem for our model.

Another interesting prediction for the future is the cutoff that we find in the luminosity of MW satellites.  We have shown that the molecular hydrogen cooling mass and the associated star formation efficiency sets a minimum luminosity for satellites that cannot be significantly dimmer than what has already been observed.  Interestingly, since the limit depends on the product $f_{\rm H_2}\,M_{\rm H_2}$, we find the same cutoff for both values of $z_{\rm H_2}$ considered.  While our theoretical value is a bit brighter than that observed to date, the difference is within the uncertainties in our calculation.  A precise measurement of this minimum luminosity in the future would be a useful test of our model, which would not be able to explain satellites much dimmer than Segue 1.

\item The luminosity of satellites between $-5>M_{\rm V}>-9$ is
dominated by stars produced through \HI\ cooling of gas before the
photoheating of the IGM at reionization.

We considered, for the first time, a model of reionization in the
MWgfr with an inside-out morphology where dense progenitors more
massive than the cooling mass, $M_4$, reionize before the rest of the
region.  This prescription is based in the potentially large optical
depth of Lyman-limit systems between the MW and the Virgo Cluster.  We
showed that this model explains the low star formation efficiency
found in previous studies that assumed instantaneous, external
reionization of the MWgfr, but both morphologies result in the same
simulated observables.

We used the observed LF of satellites to probe $\bar{z}_{\rm rei}$,
the mean reionization redshift of all regions within the MWgfr.
Depending on the choice of morphology, this represents either the
redshift at which the MWgfr is quickly ionized by the Virgo Cluster
or, since the overdensity of the IGM is small, the mean reionization
redshift of the Universe.  However, if the MWgfr does self-ionize,
then we can learn not only about the efficiency of star formation but
also about the clumpiness of the IGM.

After analyzing three possible values of $\bar{z}_{\rm rei}$ from VLII
using the inside-out reionization prescription of \S\ref{sec:HI}, we
found corresponding sets of model parameters that best fit the LF
data: $\left(\bar{z}_{\rm rei},N'_{\rm clump},f'_{\rm \HI},f_{\rm
H_2}\right)=\left\{(7.77,7,0.05,<0.001),(9.14,17,0.2,0.1),(11.2,30,1,0.4)\right\}$;
the values of $f_{\rm \HI, ex}$ for an external reionization scenario
can be calculated from equation~\ref{eq:fHIex}.  

We find that the data is most consistent with early reionization at
$\bar{z}_{\rm rei}=11.2$, consistent with the WMAP5 central value, and
very high values for the clumpiness of the IGM, star formation efficiency, and
esca	pe fraction.  This also implies a higher star formation
efficiency in molecular hydrogen cooling systems at redshifts above
$z\sim20$ than for later reionization models.
Our result also argues for early enrichment of the IGM \citep{FL03}.  
Assuming consistency with WMAP results, simulations by \citet{WC09}
have found that, for a ``normal" (i.e. non-topheavy) IMF, the product
$f_{\rm \HI}\,f_{\rm esc}$, averaged over atomic cooling halos with
virial temperatures above $8000\,{\rm K}$, to be $0.02$ with $f_{\rm
esc}\sim0.4$.  This implies $g(\delta)\sim12$ and that the cosmic IGM
is relatively smooth with only a couple recombinations per baryon and
a couple photons required, on average, to ionize each atom of
hydrogen.

However, it remains possible that the discrepancy between the data and
model predictions with later reionization results from a series of
Poisson fluctuations or cosmic variance between VLII and
the true MW history.  In such a case, we are left with three possible
points in parameter space.  These points trace out a continuous path
in the space of reionization parameters, but we were only able to
probe discrete redshifts at which VLII slices have been analyzed.
However, any external information about one of these parameters fixes
a single point that fits the data reasonably well.  For example,
evidence that stars formed in molecular hydrogen cooling halos do not
contribute to the luminosities of satellites today would argue for
reionization around $\bar{z}_{\rm rei}\approx8$.

Finally, we note that the entire MWgfr need not reionize in one
particular way; the dense progenitors could begin to self-ionize with
radiation from the Virgo Cluster completing reionization for the rest
of the region at some later time.  The results we obtained by assuming
an inside-out morphology for the entire MWgfr are valid as long as the
progenitors of the MW satellites completely reionize themselves.  This
is because the satellite data is only sensitive to what happens in
these objects.  In this case, $\bar{z}_{\rm rei}$ is the mean
reionization redshift of the Universe, since it is the time from which
the stellar mass produced by MW progenitors to reionize themselves
early is calibrated.

\item While stars formed before reionization do contribute somewhat to
the luminosity of the brightest MW satellites ($M_{\rm V}<-5$), this
population of pre-SDSS objects can be explained by a combination of
\HI\ and metal-line cooling in the most massive subhalos after
reionization.

These satellites were represented in our model by VLII objects that
have exceeded a maximum circular velocity of $V_5$ sometime in their
histories.  When we included the pre-reionization stars and
\citet{Orban08} results to determine the stellar mass formed in the
last $10\,{\rm Gyr}$, we determined that a fraction $f_5=0.01$ of the
photoheated gas retained by the largest subhalos after reionization is
converted into stars.  The majority of the gas must be kept hot or at
low density to prevent these objects from being brighter than
observed.

While old stars may be found in these satellites, they are not prime candidates for finding the relics of the first stars.

\end{enumerate}

Although we have produced some interesting results using MW satellite
data to probe the early universe, the data will become sensitive to
the parameters of even more complicated models if: (1) additional new
satellites are discovered to improve the statistics of the sample; (2)
the errors on the $M_{300}$ and the mass-to-light ratio of satellites
are improved; (3) the degree of cosmic variance between different
high-resolution MW simulations is understood; and (4) new
high-resolution simulations are developed to include baryons and
feedback processes.  Progress is already being made on some of these
fronts.  PanSTARRS \citep{Kaiser02}, the Dark Energy Survey
\citep{DESC05}, SkyMapper \citep{Keller07}, and the Large Synoptic
Survey Telescope \citep{Ivezic08} will intensively probe for
satellites populating the galactic neighborhood.  Additionally,
\citet{Ishiyama09} have compared subhalo populations for many
different simulations of galactic halos and found more scatter than
anticipated by \citet{Springel08} using a smaller sample, but their
work has not yet resolved the progenitors that formed ultra-faint
systems.  Once developed, high-resolution cosmological simulations
with baryons and feedback will test much more specific models of
reionization and open an avenue for comparisons with new observables,
in addition to those explored here, such as the half-light radius,
which promises to hold interesting clues about the high-redshift
formation physics of MW satellites.  With these improvements, our
basic methodology can be used in the future to further probe
reionization and the process of star formation in the early universe.

\section{Acknowledgements}

We thank Mike Kuhlen, Louie Strigari, Raja Guhathakurta, and Gerry
Gilmore for useful discussions.  J.D. acknowledges support from NASA
through Hubble Fellowship grant HST-HF-01194.01.  This research was
supported in part by NASA grants NNX08AL43G and LA (A.L.),
HST-AR-11268.01-A1 and NNX08AV68G (P.M.), and by Harvard University
funds.

\end{document}